\documentclass[
 reprint,
 amsmath,amssymb,
 aps,
 prl,
 longbibliography,
 noeprint
]{revtex4-2}

\usepackage{bm}              
\usepackage{graphicx}        
\usepackage{hyperref}        
\usepackage{newtxtext,newtxmath} 

\begin{document}

\title{Thermodynamic Constraints in Dynamic Random-Access Memory Cells: Experimental Verification of Energy Efficiency Limits in Information Erasure}

\author{Takase Shimizu}
\email[]{takase.shimizu@ntt.com}

\author{Kensaku Chida}

\author{Gento Yamahata}

\author{Katsuhiko Nishiguchi}

\affiliation{Basic Research Laboratories, NTT, Inc., 3-1 Morinosato-Wakamiya, Atsugi, Kanagawa 243-0198, Japan}

\begin{abstract}
We measured the energy efficiency of information erasure using silicon DRAM cells capable of counting charges on capacitors at the single-electron level.
Our measurements revealed that the efficiency decreased as the erasure error probability decreased, and notably, the Landauer limit was not achieved even under effectively infinite-time bit erasure.
By comparing the measured efficiency with the Landauer limit, we identified a thermodynamic constraint that prevents DRAM from reaching this limit: the inability to prepare the initial state in thermal equilibrium, which in turn prohibits quasistatic operations.
This finding has broad implications for DRAM cells and for many electronic circuits sharing similar structures. Furthermore, it validates our experimental approach to discovering thermodynamic constraints that impose tighter, practically relevant limits, opening a new direction in information thermodynamics research.

\end{abstract}

\maketitle

{\it Introduction}.---Information thermodynamics provides a framework for quantifying the energetic cost of information processing and clarifying its fundamental limits \cite{Landauer1961, Bennett1982, Seifert_2012, Parrondo2015, 10.1143/PTP.127.1, Wolpert_2019}.
A cornerstone is the Landauer limit, which ties the minimum heat dissipated for erasure to the entropy change of a memory, achievable through quasistatic operations \cite{Landauer1961, Bennett1982, Bérut2012, PhysRevLett.113.190601, doi:10.1126/sciadv.1501492, PhysRevLett.120.210601}.
Recent research has moved beyond quasistatic operations to explore the efficiency limits of finite-time information processing \cite{PhysRevLett.125.100602, PhysRevLett.129.120603, PhysRevX.13.011013, PhysRevLett.129.270601, PhysRevLett.127.190602, oikawa2025experimentallyachievingminimaldissipation}.

While early studies focused on device-independent limits, there is growing interest in exploring tighter and practically relevant limits that incorporate specific constraints of concrete devices \cite{PhysRevX.11.041024}.
Thermodynamic descriptions applicable to electronic circuits have recently been developed \cite{Pekola2012, PhysRevX.10.031005, PhysRevX.11.031064, PhysRevE.107.014136, PhysRevE.105.034107, PhysRevLett.129.120602, PhysRevB.109.085421, PhysRevB.106.155303, Freitas2022, yadav2025minimalthermodynamiccostcomputing, Wolpert_2020, doi:10.7566/JPSJ.92.124004}, enabling discussions on the thermodynamic efficiency of practical devices such as SRAM and inverters.
The resulting constraints are expected to guide the design of more efficient devices \cite{doi:10.1073/pnas.2321112121}, yet many remain unexplored.

Experiments are essential for identifying unknown constraints, and the first step is to compare new experimental results with established limits, followed by considering the differences.
Applying these methodologies to devices used in practical applications is crucial to finding the constraints relevant to real devices.
However, previous experiments have mainly verified known theoretical limits, such as the Landauer limit, using systems like colloidal particles, ion traps, and nanomagnetic bits \cite{Bérut2012, PhysRevLett.113.190601, PhysRevLett.120.210601, doi:10.1126/sciadv.1501492}, which are not representative of mainstream memory technologies.
Consequently, the vast efficiency gap between practical devices and the Landauer limit \cite{Zhirnov14} remains poorly understood.

\begin{figure}[tb]
\centering
\includegraphics[width=1\linewidth]{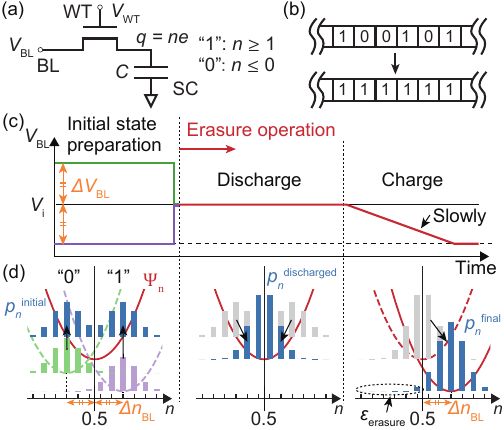}
\caption{
Experimental scheme.
(a) Circuit diagram of a DRAM cell.
(b) Conceptual diagram for the erasure of a 1-bit memory.
(c) Operational sequence of $V_{\mathrm{BL}}$ used in experiments.
(d) Calculated $p_n$ and $\Psi_n$ during the erasure process, with parameters $E_{\rm c}=8.2\ {\rm meV}$, $T=300\ {\rm K}$, and $\Delta n_{\rm BL}=2.5$.
}
\label{fig:1}
\end{figure}

Dynamic Random Access Memory (DRAM) dominates modern memory systems \cite{8976234}, but its thermodynamic performance has not been quantified.
A DRAM cell stores 1 bit in a transistor-capacitor pair (Fig.~\ref{fig:1}(a)).
Prior research has evaluated the power of an entire DRAM \cite{5695550, 10.1145/3224419} or heat dissipation of a DRAM cell during writing \cite{10231222, 10.1063/5.0152883} or erasure \cite{Orlov_2012, lent2018energy}, but has not measured the entropy change of the memory; therefore, the efficiency remains unexplored.
The primary challenge lies in the technical difficulties of simultaneously measuring entropy change of the memory and heat dissipation.

Here we measure the erasure efficiency of a silicon DRAM cell with single-electron charge sensitivity \cite{Nishiguchi_2008} under effectively infinite-time bit erasure.
We find that efficiency degrades as the erasure error is reduced, and notably, the Landauer limit was not achieved.
Our analysis shows that nonequilibrium initial states prevent quasistatic operation, imposing a device-specific constraint.
Since the circuit structure of DRAM cells commonly appears in electronic circuits, such constraints provide general insight into the energy efficiency of electronic information-processing devices.
In addition, our experimental methodology, which compares measured efficiencies with theoretical limits, provides a general approach to uncover practically relevant thermodynamic constraints and opens a new research direction in information thermodynamics.

{\it Method}.---
As in typical tests of Landauer’s principle (Fig.~\ref{fig:1}(b)), we assume a scenario where logical states ``0'' and ``1'' are written into a DRAM cell with equal probability (50\% each), and we evaluate the energy efficiency
\begin{align}
\eta = \frac{-k_B T \Delta S}{-Q}
\end{align}
by erasing the state to ``1'' and comparing the results with the Landauer limit of $\eta=1$.
Here, $\Delta S$ is the change in Shannon entropy of the logical states due to erasing, $-Q$ is the average total heat released from the system to the thermal bath for the erasure process, $k_B$ is Boltzmann's constant, and $T$ is the temperature.

A DRAM cell consists of a bitline (BL), storage capacitor (SC), and wordline transistor (WT) (Fig.~\ref{fig:1}(a)).
A 1-bit logical state is determined by the charge $q = ne$ ($e$: electron charge) on the SC (capacitance: $C$), assigned as [``0'': $n \leq 0$, ``1'': $n \geq 1$].
Here, $n$ represents the excess number of electrons on the capacitor and is an integer.
The bitline voltage $V_{\mathrm{BL}}$ controls the charging and discharging of the SC, while the WT voltage $V_{\mathrm{WT}}$ regulates the electron hopping rate between the BL and SC. 

We repeat the experimental cycle $M$ times and obtain the probability distribution $p_n$ of $n$ at each time point. Specifically, the following steps constitute a single cycle: ``initial state preparation $\rightarrow$ erasure operation''. Figure~\ref{fig:1}(c) shows the sequence of $V_{\mathrm{BL}}$ operations involved.
For initial state preparation, half of the cycles follow the green trajectory and the other half the purple one. The erasure operation is shown in red. We measure $\Delta S$ and $Q$ for the erasure operation and evaluate $\eta$.

Before detailing the experimental cycle, we define the ``erasure-to-0'' and ``erasure-to-1'' operations. Each consists of three steps: ``discharge $\rightarrow$ charge $\rightarrow$ quench''. During discharge, $V_{\rm BL}=V_{\rm i}$ is held until equilibrium. Here, $V_{\rm i} = E_{\mathrm{c}}/e$ denotes the BL voltage at which $\langle n \rangle$, the average value of $n$, takes the value 0.5 under equilibrium. $E_{\mathrm{c}} = e^2/(2C)$ is the charging energy.
For the ``erasure-to-1'' operation, during the charge step, $V_{\rm BL}$ is swept slowly from $V_{\rm i}$ to $V_{\rm i} - \Delta V_{\mathrm{BL}}$ ($\Delta V_{\mathrm{BL}}>0$), then held for equilibration.
The quench step rapidly returns $V_{\rm BL}$ to $V_{\rm i}$, so that the system is effectively erased \cite{PhysRevLett.129.270601}.
The ``erasure-to-0'' operation is defined similarly, except that $V_{\rm BL}$ is swept to $\Delta V_{\mathrm{BL}} + V_{\rm i}$ in the charge step. 

First, we prepare the initial state in which ``0'' and ``1'' are equally occupied. Assuming repeated memory use, the initial ensemble is formed from the outcomes of both ``erasure-to-0'' and ``erasure-to-1'' operations (see also the supplemental material (SM) \cite{SM}).
Practically, this is achieved by applying $V_{\mathrm{BL}} = \pm \Delta V_{\mathrm{BL}} + V_{\rm i}$ and waiting for equilibration, which corresponds to preparing the post-charge condition, and then quenching $V_{\mathrm{BL}}$ to $V_{\rm i}$.
The resulting $V_\mathrm{BL}$ trajectories are shown as the purple and green curves in Fig.~\ref{fig:1}(c).
In each branch, we attempt to write ``0'' and ``1'' $M/2$ times each.
Because of thermal fluctuations, some trials are miswritten with error probability $\varepsilon_\mathrm{init}$.
If the error probabilities for the two operations are equal, the probabilities of obtaining ``0'' and ``1'' are $1-\varepsilon_\mathrm{init}$ and $\varepsilon_\mathrm{init}$ for ``erasure-to-0'' and opposite for ``erasure-to-1.'' Averaging the two branches yields equal occupation of ``0'' and ``1'' regardless of $\varepsilon_\mathrm{init}$.

Next, we perform the ``erasure-to-1'' operation. In practice, only the discharge and charge steps are executed, as the quench step neither changes $p_n$ nor generates heat. The corresponding $V_\mathrm{BL}$ evolution is shown as the red curve in Fig.~\ref{fig:1}(c). This protocol is state-independent. We denote the erasure error probability by $\varepsilon_\mathrm{erasure}$, which is expected to satisfy $\varepsilon_\mathrm{erasure}=\varepsilon_\mathrm{init}$.

We indirectly measure the heat associated with electron hoppings between the BL and SC based on the theory proposed by Freitas et al. \cite{PhysRevX.11.031064}. In the equivalent circuit of a DRAM cell, the state function \( \Psi_n(n_{\mathrm{BL}}) \) of the system is expressed as \cite{SM}:
\begin{equation}
\label{eq:eq_state}
\Psi_n(n_{\mathrm{BL}})
= \frac{q^2}{2C}-V_{\mathrm{BL}}q
= E_c\,(n - n_{\mathrm{BL}})^2 \; - \; E_c\,n_{\mathrm{BL}}^2,
\end{equation}
assuming that the junction capacitance of WT is negligibly small compared to $C$.
Here, \( n_{\mathrm{BL}} = e\,V_{\mathrm{BL}}/(2\,E_c) \) is the normalized BL voltage. The stochastic heat associated with electron hoppings \( (n = N \rightarrow N \pm 1) \) is given by the change in the state function:
\begin{equation}
\label{eq:B}
\delta Q_{N \to N\pm1}(n_{\mathrm{BL}})
= \Psi_{N}(n_{\mathrm{BL}}) - \Psi_{N \pm 1}(n_{\mathrm{BL}}).
\end{equation}
By substituting Eq. \eqref{eq:eq_state} into \eqref{eq:B}, we obtain
\begin{equation}
       \label{eq:eq_state_explicit}
       \delta Q_\mathrm{N \rightarrow N+1}=eV_\mathrm{BL}-\mu_\mathrm{N+1},\\
       \delta Q_\mathrm{N \rightarrow N-1}=\mu_\mathrm{N}-eV_\mathrm{BL},
\end{equation}
where $\mu_N=E_cN^2-E_c(N-1)^2=2E_c(N-1/2)$ is the chemical potential of the capacitor, which corresponds to the energy change of the capacitor when $n$ increases from $N-1$ to $N$. Therefore, Eq. \eqref{eq:eq_state_explicit} represents the difference between the BL potential and the chemical potential at the moment of electron hopping---namely, the heat absorbed (or dissipated) by the electron from the heat bath during the hopping process. Thus, by measuring $n$ and $E_\mathrm{c}$, the heat associated with electron hoppings can be indirectly measured.

{\it Theory}.---We calculate \( \Delta S \) and \( Q \) for the erasure process and later compare the calculation results with experimental results.
Considering only thermal fluctuations, the probability distribution of $n$ at equilibrium is given by:
\begin{align}
p_n^{\mathrm{eq}}(n_{\mathrm{BL}})
  &= \frac{\exp\bigl[-\,\Psi_n(n_{\mathrm{BL}})/(k_B T)\bigr]}
         {\sum_{n}\exp\bigl[-\,\Psi_n(n_{\mathrm{BL}})/(k_B T)\bigr]} \label{eq:eq_pn_eq_1} \\
  &= \frac{\exp\bigl[-\,E_c\,(n - n_{\mathrm{BL}})^2/(k_B T)\bigr]}
         {\sum_{n}\exp\bigl[-\,E_c\,(n - n_{\mathrm{BL}})^2/(k_B T)\bigr]}.
\label{eq:eq_pn_eq_2}
\end{align}
This is a discrete distribution following the Gaussian shape centered around \( n_{\mathrm{BL}} \). Using \( p_n^{\mathrm{eq}}(n_{\mathrm{BL}}) \), the probability distributions immediately after quenching (initial state), after discharging, and after charging (final state) are respectively:
\begin{align}
p_n^{\mathrm{initial}}(\Delta n_{\mathrm{BL}}) &= \Bigl[p_n^{\mathrm{eq}}(-\Delta n_{\mathrm{BL}} + n_{\rm i}) + p_n^{\mathrm{eq}}(\Delta n_{\mathrm{BL}}+n_{\rm i})\Bigr]/2, \\
p_n^{\mathrm{discharged}} &= p_n^{\mathrm{eq}}(n_{\rm i}), \\
p_n^{\mathrm{final}}(\Delta n_{\mathrm{BL}}) &= p_n^{\mathrm{eq}}(\Delta n_{\mathrm{BL}}+n_{\rm i}),
\end{align}
as plotted in Fig.~\ref{fig:1}(d), where $\Delta n_{\mathrm{BL}} = e\,\Delta V_{\mathrm{BL}}/(2\,E_c)$.
Thus, the initial distribution \( p_n^{\mathrm{initial}} \) is a nonequilibrium state composed of two overlapping Gaussian shape distributions centered around \( n=\pm \Delta n_{\rm BL}+n_{\rm i} \), with \( n_{\mathrm{i}} = e\,V_{\mathrm{i}}/(2\,E_c) = 0.5 \).
As discussed later, this initial state precludes quasistatic operation in DRAM cells.
Now, the entropy change is calculated as \( \Delta S = S\bigl(p^{\mathrm{final}}_n\bigr) - S\bigl(p^{\mathrm{initial}}_n\bigr) \), where $S$ is the Shannon entropy of the logical states.

Let the average heat for the discharge and charge processes be \( Q_{\mathrm{discharge}} \) and \( Q_{\mathrm{charge}} \), respectively. The average total heat \( Q \) is given by \( Q = Q_{\mathrm{discharge}} + Q_{\mathrm{charge}} \), with each component calculated as:
\begin{equation}
\label{eq:Q_disc}
Q_{\mathrm{discharge}}(\Delta n_{\mathrm{BL}})
= \sum_{n} \Bigl[p_n^{\mathrm{discharged}} - p_n^{\mathrm{initial}}\Bigr]\,\Psi_n(n_{\rm i}),
\end{equation}
\begin{equation}
\label{eq:A}
Q_{\mathrm{charge}}(\Delta n_{\mathrm{BL}})
= k_{\rm B}T\left( s\bigl(p^{\mathrm{final}}\bigr) - s\bigl(p^{\mathrm{discharged}}\bigr) \right),
\end{equation}
where \( s(p) \equiv - \sum_n p_n \ln p_n \) is the Shannon entropy with respect to \( n \).
In Eq.~\eqref{eq:Q_disc}, since the discharge process occurs at fixed \( n_{\mathrm{BL}} = n_{\rm i} \), the heat is calculated from the state function differences weighted by probability changes.
As for Eq.~\eqref{eq:A}, the charging process is assumed to be quasistatic, resulting in heat that equals the change in entropy.
\begin{figure}[tb]
       \centering
       \includegraphics[width=1\linewidth]{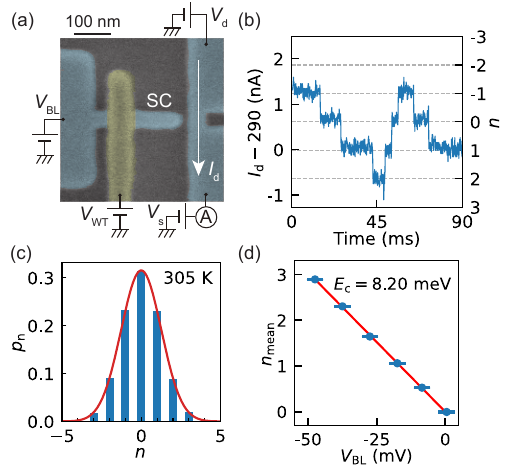}
       \caption{%
       Experimental setup and characterization.
       (a) Scanning electron micrograph of a device with measurement setup.
       (b) The sensor transistor current ($I_{\rm d}$) under equilibrium conditions.
       (c) $p_n$ at $n_{\rm BL}=0$, with a Gaussian fit (red line).
       (d) The mean value of $n$ ($n_{\rm mean}$) obtained from Gaussian fit of the distribution of $n$ at various $V_{\rm BL}$. The error bars for $n_{\rm mean}$ indicate the standard error of the mean.
       }
       \label{fig:2}
       \end{figure}

{\it Setup and measurement of $E_{\rm c}$ and $T$}.---To accurately measure $Q$ and $\Delta S$, we used a silicon DRAM cell device capable of measuring the time evolution of $n$ from the current $I_{\rm d}$ of a sensor transistor adjacent to the SC, as shown in Fig.~\ref{fig:2}(a) \cite{Nishiguchi_2008}.
We focus on the DRAM cell in Fig. 1(a) and do not consider the readout cost of the sensor \cite{SM, 5rtj-djfk}.
All measurements were conducted with the sample fixed on a stage thermally controlled at 300 K. 
Throughout the experiment, $V_\mathrm{WT}$ is kept constant at a value that ensures the hopping rate is sufficiently slow to allow individual electron hopping events to be measured.
We modeled the system using the equivalent DRAM circuit in Fig.~\ref{fig:1}(a) and evaluated $Q$ using Eq.~\eqref{eq:B}. 
In practice, the SC is also capacitively coupled to the WT, upper gate, back gate, and sensor transistor, but these voltages were held constant and only result in an offset in $V_{\rm BL}$, leaving our circuit assumption valid \cite{SM}.

First, we conducted experiments to determine $E_{\rm c}$ and $T$.
Figure \ref{fig:2}(b) shows an example of the time evolution of $I_{\rm d}$ under equilibrium conditions.
The discrete changes in $I_{\rm d}$ correspond to changes in $n$.
Figure \ref{fig:2}(c) presents $p_n$ at $n_{\rm BL}=0$, which is well fitted by a Gaussian function.
Figure \ref{fig:2}(d) shows the mean value of $n$ ($n_{\rm mean}$) obtained from Gaussian fit of $p_n$ at various $V_{\rm BL}$. 
According to Eq.~\eqref{eq:eq_pn_eq_2}, $n_{\rm mean}$ equals $n_{\rm BL}$ at equilibrium.
Thus, $E_{\rm c}$ necessary to determine $\Psi_n$ can be extracted from the slope $d n_{\rm mean}/dV_{\rm BL}$:
\begin{equation}
E_{\rm c} = \frac{e}{2}\left(\frac{d n_{\rm mean}}{dV_{\rm BL}}\right)^{-1} = 8.20\pm0.08\,\mathrm{meV}.
\end{equation}
The reference point $V_{\rm BL}=0$ was chosen such that $n_{\rm mean}=0$.
The temperature derived from the variance of the Gaussian fit in Fig. \ref{fig:2}(c) was calculated to be $T = 305\pm 5\,\mathrm{K}$, only 1.7\% higher than the stage temperature of 300 K.
This consistency confirms the assumed equilibrium state, indicates that external noise and sensor-current-induced Joule heating \cite{10.1063/5.0114584} are negligible, and supports the validity of the measurement.
The assumption that the junction capacitance is negligible compared with $C$ is also verified \cite{SM}.
For subsequent analysis, we used $T = 305\ {\rm K}$.

\begin{figure}[tb]
\centering
\includegraphics[width=1\linewidth]{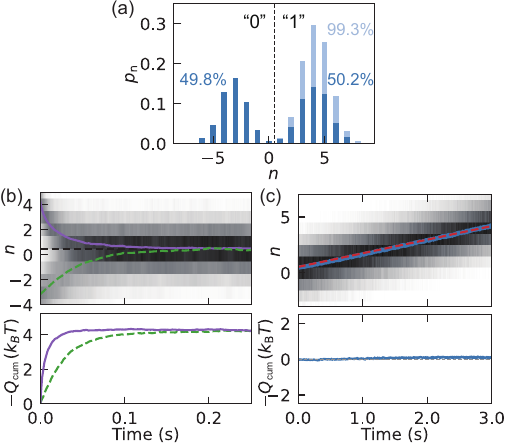}
\caption{%
Erasure process at $\Delta n_{\rm BL}= 3.7$.
(a) $p_n$ before (dark blue) and after (light blue) erasure, with logical-state occupations.
(b) Top: $\left <n \right >$ during discharge. The green dashed and purple solid lines represent cases starting immediately after the initial state preparation with the quench from $V_\mathrm{BL}=V_\mathrm{i} \mp \Delta V_\mathrm{BL}$, respectively. The grayscale background indicates $p_n$. Bottom: Ensemble-averaged cumulative heat dissipation $-Q_\mathrm{cum}$ during discharge.
(c) Top: $p_n$ during charging, with $\langle n\rangle$ (blue) and $n_\mathrm{BL}$ (red).
Bottom: $-Q_\mathrm{cum}$ during charging.
}
\label{fig:3}
\end{figure}

{\it Time evolution during erasure}.---We repeated the erasure cycle $M=1998$ times, and here we demonstrate the system behavior during the erasure process with $\Delta n_{\mathrm{BL}} = 3.7$.
Figure~\ref{fig:3}(a) shows $p_n$ before erasure (dark blue), where ``0'' and ``1'' are nearly equally occupied, confirming the initial state preparation in Fig.~\ref{fig:1}(d).
Fig.~\ref{fig:3}(b) (top) presents the time evolution of $p_n$ during discharge.
The green dashed and purple solid lines represent $\langle n \rangle$ after quenches from  $V_\mathrm{BL}=V_\mathrm{i} \pm \Delta V_\mathrm{BL}$. 
$n_{\rm mean}$ is not shown here because the distribution is not necessarily a Gaussian shape.
Both trajectories relax toward $\langle n\rangle=0.5$, with the faster purple relaxation reflecting the asymmetry of the hopping rates with respect to $n$ \cite{Nishiguchi_2014}. 
The bottom panel shows ensemble-averaged cumulative heat dissipation $-Q_\mathrm{cum}$ during discharge, obtained from Eq. \eqref{eq:eq_state_explicit}.
Although the initial rise differs due to energy-dependent discharge speed, the final values converge, indicating that total heat generated during discharge is independent of discharge speed, consistent with Eq.~\eqref{eq:Q_disc}.
In Fig.~\ref{fig:3}(c) (top), the gray plot shows $p_n$ during charging, with $\langle n\rangle$ (blue) closely following $n_{\mathrm{BL}}$ (red), indicating thermal equilibrium and a quasistatic process. 
The bottom panel shows that $-Q_\mathrm{cum}$ during charging is minimal, owing to the small change in the distribution and thus in entropy of the system.
The final-state histogram in Fig.~\ref{fig:3}(a) (light blue) confirms that the erasure operation has been performed.

\begin{figure}[tb]
\centering
\includegraphics[width=1\linewidth]{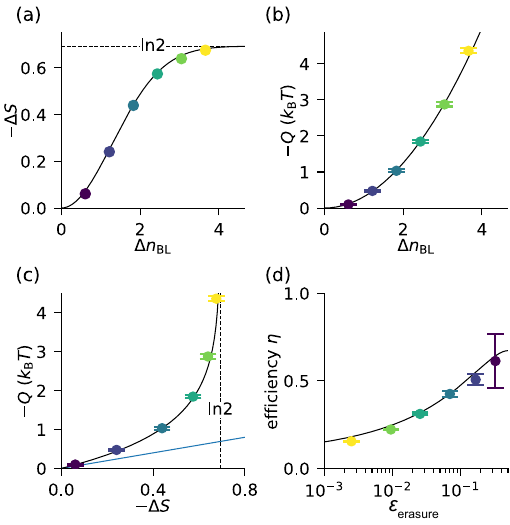}
\caption{%
Erasure experiment results. Time evolution data are in the SM \cite{SM}.
(a) $-\Delta S$ versus $\Delta n_{\mathrm{BL}}$.
(b) $-Q$ versus $\Delta n_{\mathrm{BL}}$.
(c) $-Q$ versus $-\Delta S$. Blue line: The Landauer limit ($Q/k_{\rm B}T=\Delta S$).
(d) Efficiency $\eta$ versus erasure error probability $\varepsilon_\mathrm{erasure}$.
Black solid lines: theoretical calculations. Error bars: standard error of the mean for $Q$.
}
\label{fig:4}
\end{figure}

{\it Erasure results}.---Figure~\ref{fig:4}(a) shows that $-\Delta S$ increases monotonically with $\Delta n_{\mathrm{BL}}$ and approaches $\ln 2$, reflecting that $S$ of the initial state is always $\ln 2$, while $S$ of the final state decreases and approaches zero once $\Delta n_{\mathrm{BL}}$ exceeds thermal fluctuations of $n$. 
Figure~\ref{fig:4}(b) shows that $-Q$ also increases with $\Delta n_{\mathrm{BL}}$, and Fig.~\ref{fig:4}(c) indicates that $-Q$ diverges as $-\Delta S \rightarrow \ln 2$. 
The Landauer limit $Q/k_{\rm B}T = \Delta S$ (blue line) lies below all measured values, demonstrating that the heat dissipation always exceeds the Landauer limit.
Figure~\ref{fig:4}(d) shows $\eta$, which remains below unity for all $\varepsilon_\mathrm{erasure}$ and decreases as $\varepsilon_\mathrm{erasure}$ decreases.
The solid curves in Fig.~\ref{fig:4}(a--d) show theoretical predictions that agree well with experiment, confirming the validity of the model in Fig.~\ref{fig:1}(d). 
Overall, the results show that this DRAM cell cannot reach the Landauer limit.

{\it Discussion}.---
The DRAM cell did not reach the Landauer limit because its initial state is nonequilibrium. Systems that can achieve the Landauer limit, like double-well potentials or two-level systems, start in equilibrium and thus allow quasistatic operation. In contrast, the DRAM's nonequilibrium initial state makes the discharge process non-quasistatic, leading to substantial heat dissipation (Fig.~\ref{fig:3}(b)).

Regardless of microscopic details, the DRAM circuit in Fig.~\ref{fig:1}(a) cannot perform quasistatic 1-bit erasure.
No choice of the controllable parameters $V_\mathrm{BL}$ and $V_\mathrm{WT}$ yields a state function whose equilibrium distribution matches $p_\mathrm{init}$.  
Such a distribution would require a double-well state function with minima at $0.5\pm \Delta n_\mathrm{BL}$.  
However, Eq. \eqref{eq:eq_state} shows that changing $V_\mathrm{BL}$ only shifts a single-well parabola, and $V_\mathrm{WT}$ affects only hopping rates, not the state function.
Thus, DRAM cells cannot prepare their initial state in equilibrium, preventing the quasistatic operations required to reach the Landauer limit. This thermodynamic constraint highlights a fundamental difference from two-level or double-well potential systems \cite{SM}, offering a thermodynamic interpretation of volatile memory.

Finite-time operations during the evolution of the probability distribution are often regarded as a primary reason for failing to reach the Landauer limit. However, here we show that the Landauer limit is not reached even under effectively infinite-time operations of the discharge and charge processes governing this evolution. Note that the final quench only erases the writing history, marking the completion of the operation. Since the probability distribution is unchanged, it does not contribute to heat dissipation during the erasure itself.

However, the quench can cause heat dissipation in the subsequent erasure cycle. When restricted to the DRAM cell, attributing the extra heat above the Landauer limit to the initial nonequilibrium distribution is effectively equivalent to attributing it to the preceding quench, which generates that nonequilibrium state. Heat generation associated with such quenches has been qualitatively discussed in Refs.~\cite{10231222, 10.1063/5.0152883, Orlov_2012, lent2018energy}. By contrast, in a two-level system, equilibrium can still be established after the quench \cite{SM}. Thus, from a thermodynamically universal perspective beyond the DRAM cell, the quench is only a possible contributing factor, whereas the initial nonequilibrium distribution is sufficient to prevent reaching the Landauer limit.

The initial nonequilibrium distribution arises not from a single operation but from the ensemble of outcomes of both “erase-to-0” and “erase-to-1” operations during information processing---a characteristic feature of information thermodynamics.
Its origin likely lies in the system being a multi-level system (three or more states) with a single potential well; however, a detailed investigation of this aspect is left for future studies.

{\it Conclusion and outlook}.---
We measured the energy efficiency of information erasure in DRAM cells and found that they do not reach the Landauer limit even under effectively infinite-time bit erasure.
The key constraint is that the initial state cannot be prepared in thermal equilibrium, a distinctive property of DRAM cells.
As most information-processing circuits are built from transistors and capacitors, and DRAM cells represent a minimal such unit, this constraint likely applies broadly across modern electronics.

This study suggests several directions for future research.
First, exploring the dependence of erasure efficiency on $E_c$ and $k_B T$ is intriguing, particularly in regimes $E_c / (k_B T) \gg 1$, where the system may approach two-level behavior.
Second, the efficiency of finite-time operations \cite{PhysRevLett.125.100602, PhysRevLett.127.190602, PhysRevLett.129.120603, PhysRevLett.129.270601, PhysRevX.13.011013, oikawa2025experimentallyachievingminimaldissipation} and multi-bit DRAM cell architectures \cite{10.1063/1.2200475} should be investigated.
Third, studying the reliability of the readout process, where sense amplifiers in actual DRAM extract free energy from nonequilibrium states to determine stored bits, is intriguing, especially for studying Maxwell's demon-like more efficient memory operation.
Fourth, this methodology can be applied to other electronic circuits, such as SRAM cells \cite{PhysRevX.11.031064}, by integrating the sensor transistor adjacent to the free conductors.
Finally, this work motivates a new classification framework for information-processing devices based on the thermodynamic constraints that determine their fundamental efficiency limits. 

We thank Tan Van Vu, Keiji Saito, Naruo Ohga, Yoshihiko Hasegawa, Mart\'i Perarnau-Llobet, Motoki Asano, and Yasuhiro Tokura for fruitful discussions.

\bibliography{references}

\onecolumngrid
\clearpage
\appendix

\renewcommand{\thesection}{S\arabic{section}}
\renewcommand{\theequation}{S\arabic{equation}}
\renewcommand{\thefigure}{S\arabic{figure}}
\setcounter{equation}{0}
\setcounter{figure}{0}

\begin{center}
       \textbf{\large Supplemental Material for ``Thermodynamic Constraints in Dynamic Random-Access Memory Cells: Experimental Verification of Energy Efficiency Limits in Information Erasure''}
\end{center}
\vspace{2em}

\section{Device structure and fabrication process}
The device structure is mainly the same as the one used in Ref.~\cite{Nishiguchi_2014}. Wire channels were defined in a 13-nm-thick silicon-on-insulator layer and thermally oxidized to form the gate oxide. The WT gate, consisting of phosphorus-doped polycrystalline Si (poly-Si), was formed on the wire channel. An interlayer oxide and an upper poly-Si gate were then deposited and patterned to cover the entire region. The substrate (wafer backside) was biased and used as a global back gate.

\section{Details of the measurement method}
$V_\mathrm{BL}$, $V_\mathrm{WL}$, and the source-drain voltage of the sensor transistor were applied using an arbitrary waveform generator (Zurich Instruments HDAWG).
A low-pass filter with a cutoff frequency of 1.9~MHz (Mini-Circuits BLP-1.9+) was inserted between the arbitrary waveform generator and the sample.
The sensor current was converted to voltage using an NF Corporation CA5351 with a filter setting of $100~\mathrm{\mu s}$, and its output voltage was measured using a National Instruments digitizer (PXIe-5172).

The measurement time step was $100~\mathrm{\mu s}$, which is much longer than the characteristic voltage change time of the BL during the quench process (approximately $0.2~\mathrm{\mu s}$, corresponding to the 90\% rise time). Therefore, the effect of capacitive cross-talk when $V_\mathrm{BL}$ changes abruptly is negligible.

During the measurement, the upper gate voltage was set to $1.427~\mathrm{V}$, the WT voltage to $-0.873~\mathrm{V}$, and the back gate voltage to $0.427~\mathrm{V}$. A source-drain voltage of $0.2~\mathrm{V}$ ($V_\mathrm{d}=0.627~\mathrm{V}$, $V_\mathrm{s}=0.427~\mathrm{V}$) was applied to the sensor transistor.

\section{Derivation of the State Function $\Psi(q)$}\label{sec:derivation_of_state_function}

\begin{figure}[tb]
  \centering
  \includegraphics[width=0.3\linewidth]{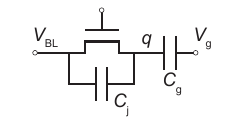}
  \caption{%
  Circuit diagram for the general calculation.
  }
  \label{fig:S1}
  \end{figure}

We derive the state function $\Psi(q)$ used in the main text. Our approach follows the general framework for nonlinear electronic circuits introduced by Freitas\,\textit{et\,al.}\,\cite{PhysRevX.11.031064}. In this framework, when an isothermal and closed circuit is opened by connecting some or all components to a voltage source, the energy change during an electron transition can be expressed as the change in the state function (see also [Eq.~(78) of Ref.~\cite{PhysRevX.11.031064}]):
\begin{equation}\label{eq:psi_general}
  \Psi(\bm{q}) = \Phi(\bm{q}) - \sum_{n_{p}} V_{n_{p}}\,L_{n_{p}}(\bm{q}),
\end{equation}
where $\bm{q}$ represents the charges on all \emph{free} conductors, $V_{n_{p}}$ is the voltage of the \emph{regulated} conductor $n_{p}$, and $L_{n_{p}}(\bm{q})$ is the total free charge in the connected component to which conductor $n_{p}$ belongs. Conductors with fixed potentials are referred to as regulated conductors, while the rest are free conductors.
The function $\Phi(\bm{q})$ in Eq.~\eqref{eq:psi_general} is generally given in Ref.~\cite{PhysRevX.11.031064} by
\begin{equation}\label{eq:phi_general}
  \Phi(\bm{q}) = E(\bm{q}) - \bm{V}_r^{\mathrm T} \bm{C}_m^{\mathrm T} \bm{C}^{-1} \bm{q},
\end{equation}
where 
\begin{equation}
E(\bm{q}) = \frac{1}{2}\,\bm{q}^{\mathsf{T}}\,\bm{C}^{-1}\bm{q} + \frac{1}{2}\,\bm{V}_r^{\mathsf{T}}\left(\bm{C}_r - \bm{C}_m^{\mathsf{T}}\,\bm{C}^{-1}\bm{C}_m\right)\bm{V}_r
\end{equation}
is the total electrostatic energy.
Here, $\bm{C}$ is the capacitance matrix of the free conductors, $\bm{C}_r$ is the matrix of the regulated conductors, $\bm{C}_m$ is the matrix with the mutual capacitances between free and regulated conductors, and $\bm{V}_r$ is the vector of regulated voltages.
For generality, we consider the circuit depicted in Fig.~\ref{fig:S1}. In this setup, the SC is capacitively coupled to the regulated conductor with a voltage $V_{\rm g}$, serving as the environmental gate voltage (such as the upper gate voltage) as mentioned in the main text.

We assume a capacitive coupling $C_{\rm j}$ exists between BL and SC. To model the DRAM cell, we set $V_g=0$ later. 
To obtain each matrix, we consider the relation between the charges and voltages in the system:
\begin{align}
  \begin{bmatrix}
  q \\
  q_{\rm BL} \\
  q_{\rm g} \\
  \end{bmatrix}
  =
  \begin{bmatrix}
  C_{\rm j} + C_{\rm g} & \vline & -C_{\rm j} & -C_{\rm g} \\
  \hline
  -C_{\rm j} & \vline & C_{\rm j} & 0 \\
  -C_{\rm g} & \vline & 0 & C_{\rm g} \\
  \end{bmatrix}
  \begin{bmatrix}
  V \\
  V_{\rm BL} \\
  V_{\rm g} \\
  \end{bmatrix},
\end{align}
and thus, each matrix is given by
\begin{align}
  \bm{C} = 
  \begin{bmatrix}
  C_{\rm j} + C_{\rm g}
  \end{bmatrix}, \quad
  \bm{C}_{\rm m} =
  \begin{bmatrix}
  -C_{\rm j} & -C_{\rm g}
  \end{bmatrix}, \quad
  \bm{C}_{\rm r} =
  \begin{bmatrix}
  C_{\rm j} & 0 \\
  0 & C_{\rm g}
  \end{bmatrix}
\end{align}
Regarding the second term in Eq.~\eqref{eq:psi_general}, the only connected component to which the free conductor belongs is BL, and thus it is given by
\begin{equation}\label{eq:single_regulated}
  \sum_{n_{p}} V_{n_{p}}\,L_{n_{p}}(\bm{q}) = V_{\mathrm{BL}} q.
\end{equation}
Then, with $\bm{q} = \begin{bmatrix} q \end{bmatrix}$ and by taking $C_\Sigma = C_{\rm j} + C_{\rm g}$, we obtain the state function of the circuit of Fig.~\ref{fig:S1} as
\begin{align}
  \Psi(\bm{q})
  &=
  \frac{q^2}{2C_\Sigma}
  +
  \frac{(V_{\rm g}-V_{\rm BL})C_{\rm g}}{C_\Sigma}q
  +
  \frac{C_{\rm j}C_{\rm g}(V_{\rm BL}-V_{\rm g})^2}{2C_\Sigma}\\
  &=
  \frac{1}{2C_\Sigma}
  \Bigl( q - C_g\!\left( V_{\mathrm{BL}} - V_g \right) \Bigr)^2
  + \frac{C_g\!\left( C_j - C_g \right)}{2C_\Sigma}
    \left( V_{\mathrm{BL}} - V_g \right)^2
\end{align}
As is evident from this equation, the gate voltage $V_{\rm g}$ for the SC acts merely as an offset to $V_{\rm BL}$.

To consider the DRAM cell depicted in Fig. 1(a) in the main text, we set $V_{\rm g}=0$, and then we obtain
\begin{align}\label{eq:phi_general_expanded}
  \Psi(q) = \frac{q^2}{2C_\Sigma}
  -
  V_{\rm BL}\left(1 - \frac{C_{\rm j}}{C_\Sigma}\right)q
  +
  \frac{C_{\rm j}C_{\rm g}V_{\rm BL}^2}{2C_\Sigma}
\end{align}
Assuming $C_{\rm j}\ll C_{\rm g}$ and neglecting the $q$-independent third term, which does not influence Eq. (3) in the main text, we can approximate Eq.~\eqref{eq:phi_general_expanded} as follows:
\begin{align}
\Psi(q) &= \frac{q^2}{2C_g} - V_{\mathrm{BL}}q,
\end{align}
This is the equation referenced in the main text as Eq. (2).

Let us discuss the validity of the approximation $C_{\rm j}\ll C_{\rm g}$.
By taking $q=ne$ and by completing the square of Eq.~\eqref{eq:phi_general_expanded}, we obtain
\begin{align}
  \Psi(n) &= \frac{e^2}{2C_\Sigma}\left(n - \left(1-\frac{C_{\rm j}}{C_\Sigma}\right)\frac{C_\Sigma V_{\mathrm{BL}}}{e}\right)^2 - \frac{e^2}{2C_\Sigma}\left(1-\frac{C_{\rm j}}{C_\Sigma}\right)^2\frac{C_\Sigma^2V_{\mathrm{BL}}^2}{e^2} + \frac{1}{2}\left(C_{\rm r}-\frac{C_{\rm j}^2}{C_\Sigma}\right)V_{\mathrm{BL}}^2\\
  &= E_{\rm c}\left(n - \left(1-\frac{C_{\rm j}}{C_\Sigma}\right)\frac{C_\Sigma V_{\mathrm{BL}}}{e}\right)^2 + \text{const.},
\end{align}
where $E_{\rm c} = e^2/(2C_\Sigma)$.
Following Eq. (4) in the main text, we obtain
\begin{align}
n_{\rm mean} &= \left(1-\frac{C_{\rm j}}{C_\Sigma}\right)\frac{C_\Sigma V_{\mathrm{BL}}}{e}.
\end{align}
Then, without the approximation of $C_{\rm j}\ll C_{\rm g}$, we obtain
\begin{align}
  E_{\rm c} = \frac{e}{2}\left(\frac{dn_{\rm mean}}{dV_{\rm BL}}\right)^{-1}\left(1-\frac{C_{\rm j}}{C_\Sigma}\right).
\end{align}
Here, let us express the relation used for experimentally obtained charging energy in the main text as
\begin{align}\label{eq:E_c_exp}
  E_{\rm c}^{\rm exp} = \frac{e}{2}\left(\frac{dn_{\rm mean}}{dV_{\rm BL}}\right)^{-1}
\end{align}
Thus, the actual charging energy should be obtained as
\begin{align}
  E_{\rm c} = E_{\rm c}^{\rm exp}\left(1-\frac{C_{\rm j}}{C_\Sigma}\right),
\end{align}
showing that the charging energy is overestimated by a factor of $(1-C_{\rm j}/C_\Sigma)^{-1}$ when $C_{\rm j}$ is ignored.
Let us assume that the true temperature of the sample is $T=300\,{\rm K}$, and the difference between this and the experimentally obtained temperature $T^{\rm exp}=305\,{\rm K}$ is attributable to the difference between $E_{\rm c}^{\rm exp}$ and $E_{\rm c}$.
Using the variance $\sigma^2$ of the equilibrium distribution, we have
$k_{\rm B}T = 2\sigma^2 E_{\rm c}$ and $k_{\rm B}T^{\rm exp} = 2\sigma^2 E_{\rm c}^{\rm exp}$, and thus $T/T^{\rm exp} = E_{\rm c}/E_{\rm c}^{\rm exp}$.
Comparing these, we obtain
\begin{align}
  \frac{T}{T^{\rm exp}} = \frac{E_{\rm c}}{E_{\rm c}^{\rm exp}} = \left(1-\frac{C_{\rm j}}{C_\Sigma}\right).
\end{align}
Solving with respect to $C_{\rm j}/C_{\rm g}$, we obtain
\begin{align}
  \frac{C_{\rm j}}{C_{\rm g}} = \frac{T^{\rm exp}}{T} -1 = \frac{305}{300} - 1 = 1.7 \%.
\end{align}
Thus, the validity of the approximation $C_{\rm j}\ll C_{\rm g}$ is confirmed.

The estimated junction capacitance $C_{\rm j}=0.017C_\mathrm{g}=0.17~\mathrm{aF}$ is reasonable. This small $C_{\rm j}$ is achieved by applying a sufficiently negative $V_\mathrm{WL}$ to suppress the electron hopping rate, allowing changes in $n$ to be detected. As $V_\mathrm{WL}$ is made more negative, the separation between electrons in the BL and SC increases, leading to a reduced $C_{\rm j}$. To substantiate this explanation, we can estimate $C_\mathrm{j}$ by examining the device's geometry. The nanowire's width and height are approximately 30 nm and 13 nm, respectively, and we assume a 150 nm distance between the BL and SC electrons. The capacitance, estimated using a parallel plate capacitor model that includes the silicon, is calculated to be $0.27~\mathrm{aF}$. The WT gate screens the electrostatic coupling between the BL and SC, so the actual junction capacitance is expected to be smaller, which is consistent with the experimentally extracted value of $0.17~\mathrm{aF}$.

\newpage
\section{Time Evolution of $S$, $\Delta S$, and $Q_\mathrm{cum}$ during the erasure process}

\begin{figure}[htbp]
  \centering
  \includegraphics[width=1\linewidth]{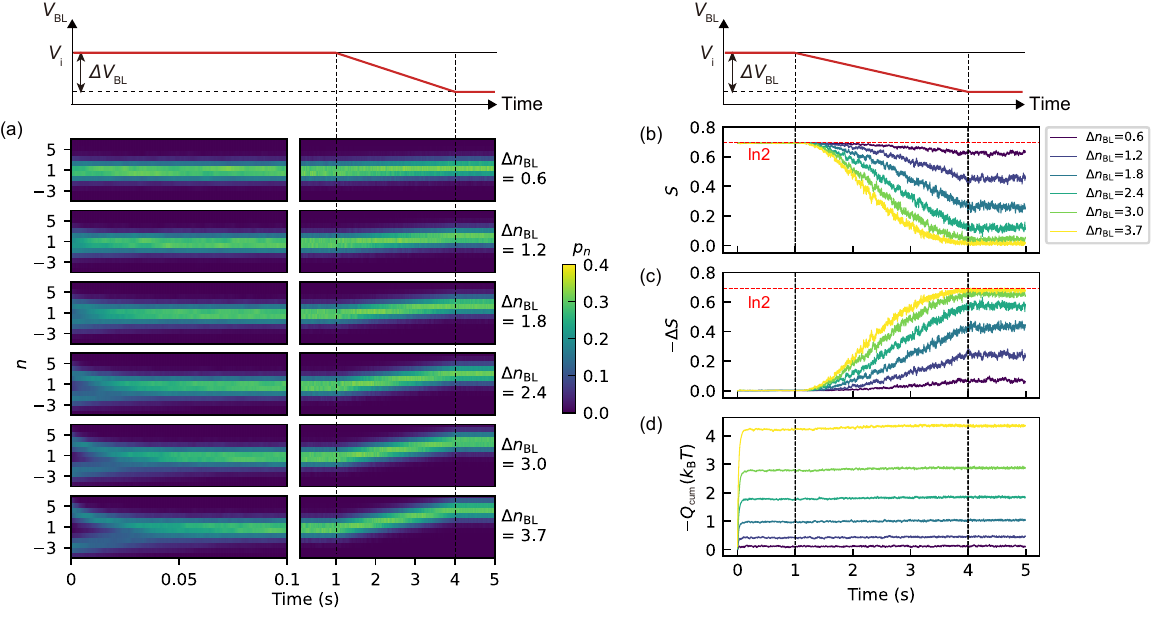}
  \caption{Time evolution of (a) $p_n$, (b) $S$, (c) $-\Delta S$, and (d) $-Q_\mathrm{cum}$ during the erasure process for several values of $\Delta n_\mathrm{BL}$.}
  \label{fig:Suppl_TimeEvol}
\end{figure}

Figures~\ref{fig:Suppl_TimeEvol}(a–d) show the time evolution of various quantities during the erasure operation, which are used to plot the points in Fig. 4 in the main text. The schematic diagram at the top of each graph represents the time evolution of $V_\mathrm{BL}$.
Figure~\ref{fig:Suppl_TimeEvol}(a) shows the time evolution of $p_n$. For clarity, the time period from 0 to 0.1 s is magnified.
As $\Delta n_\mathrm{BL}$ increases, the two distributions at the initial state (time = 0~s) become more separated. Figure~\ref{fig:Suppl_TimeEvol}(b) shows the time evolution of the logical-state entropy $S$. At time = 0~s, $S = \ln 2$ regardless of $\Delta n_\mathrm{BL}$, indicating that the logical states are written with equal probability (50\% each). Until the discharge is completed at time = 1~s, $S$ remains at $\ln 2$. 
This means that although relaxation occurs during discharge, the logical-state distribution remains unchanged and symmetric, reflecting the symmetry of $p_n^\mathrm{discharged}$ for the two logical states. (We note that although the asymmetry of the transition rate with respect to $n$ may cause a slight bias in the logical state distribution during discharge, this effect is negligibly small in the present parameter regime.)
After that, as $V_\mathrm{BL}$ is swept toward $V_\mathrm{i} - \Delta V_\mathrm{BL}$, the probability of the logical ``1'' state increases, causing $S$ to decrease. Correspondingly, $-\Delta S$ increases, as shown in Fig.~\ref{fig:Suppl_TimeEvol}(c). Figure~\ref{fig:Suppl_TimeEvol}(d) shows the time evolution of $-Q_\mathrm{cum}$. After the main heat dissipation occurs during the discharge process, almost no additional heat is produced. $-Q_\mathrm{cum}$ increases with increasing $\Delta n_\mathrm{BL}$.

\section{Comparison between the DRAM cells and two-level systems}
The essential difference between our experiment and previous studies is that we are forced to incorporate a relaxation process into our protocol. This requirement arises because the initial state is out of equilibrium, which prohibits quasi-static operations. To clarify this point, we compare information erasure in a two-level system and a DRAM cell.

\begin{figure}[htbp]
  \centering
  \includegraphics[width=0.7\linewidth]{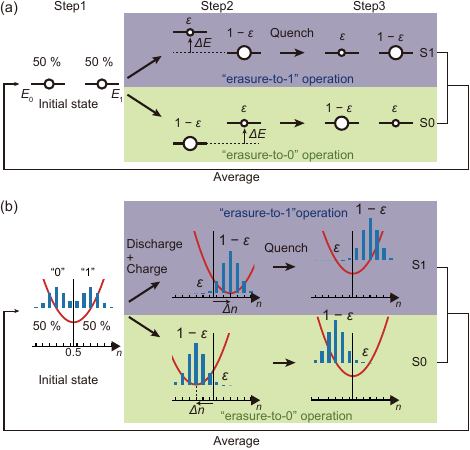}
  \caption{Schematic diagram of the erasure process. (a) Two-level system. (b) DRAM cell.}
  \label{fig:Suppl_Comparison_TwoLevel_DRAMCell}
\end{figure}

Figure \ref{fig:Suppl_Comparison_TwoLevel_DRAMCell}(a) shows an erasure process for a two-level system labeled 0 and 1, each coupled to a single heat bath. Let the corresponding energies be $E_0$ and $E_1$. 
    
\begin{enumerate}
\item[] Step 1: Setting $E_0=E_1=0$, we can prepare the initial distribution $p_0=p_1=1/2$ as a thermal equilibrium state.  
\item[] Step 2: We start the erasure operation to state "1". By increasing $E_0$ by $\Delta E$, we obtain $p_0=\epsilon$ and $p_1=1-\epsilon$. Here, $\epsilon$ is the error probability.  
\item[] Step 3: Subsequently, quenching $E_0$ back to $E_0=0$ effectively erases the system. We name this final state S1.  
For erasure to "0", $E_1$ is instead increased by $\Delta E$ at Step 2, resulting in $p_0=1-\epsilon$ and $p_1=\epsilon$; after quenching, we obtain the final state S0.
\end{enumerate}

Crucially, the ensemble average of S0 and S1 reproduces the same equilibrium distribution as prepared in Step 1. Thus, while the initial ensemble should in principle be prepared as an ensemble of S0 and S1, it coincides with the equilibrium distribution obtained with $E_0=E_1$, and conventional experiments therefore use the equilibrium state without explicitly performing such preparation.

In contrast, in our DRAM-cell experiment, the ensemble composed of S0 and S1 becomes a nonequilibrium state, which constitutes the explicit difference from previous studies. For clarity, we restate the procedure here (partly overlapping with the main text):
\begin{enumerate}
\item[] Step 1: The initial state is given by $p_n^\mathrm{init}$, where the logical states "0" and "1" are each occupied with 50\%. The state function is $\Psi_n(0.5)$.  
\item[] Step 2: After the discharge and charge operation, the system reaches equilibrium distributions $p_n^\mathrm{eq}(0.5\pm \Delta n_\mathrm{BL})$ at state functions $\Psi_n(0.5\pm \Delta n_\mathrm{BL})$.  
\item[] Step 3: The system is then quenched back to $\Psi_n(0.5)$, completing the erasure operation.
\end{enumerate}

The ensemble prepared with equal numbers of such operations (``erasure-to-1'' and ``erasure-to-0'') constitutes the initial state in our experiment, which is clearly a nonequilibrium distribution. For simplicity, we experimentally used the ensemble of equilibrium states corresponding to $V_\mathrm{BL}=V_i \pm \Delta V_\mathrm{BL}$ as the initial state. Its essential meaning, however, is that it represents the ensemble composed of the results of ``erasure-to-0'' and ``erasure-to-1'' operations. The nonequilibrium nature of this ensemble is precisely the fundamental reason why the Landauer limit cannot be reached in our experiment, and this constitutes the essential difference from conventional two-level systems.

\section{Discussion on the readout cost of the charge sensor}
In some studies, considering the readout cost of charge sensors is important. For example, in Ref. \cite{5rtj-djfk}, the charge sensor that reads out the clock information is regarded as part of the system, and the thermodynamic cost of the readout is explicitly taken into account. In particular, the paper resolves an apparent paradox: an operation that is intrinsically irreversible (the clock operation) can seem to be realized with zero entropy production if the clock information is continuously read out by a charge sensor. This paradox is resolved by focusing on the readout cost. These considerations indicate that incorporating the readout cost is essential for correctly analyzing clocks.
 
In contrast, in our experiment the charge sensor functions solely as a probe to measure heat dissipation and entropy changes during the erasure operation, and it is not an essential element of the erasure itself. Our study focuses exclusively on the DRAM cell shown in Fig. 1(a) of the main text; therefore, we do not include the readout cost of the charge sensor. Unlike Szilard engines and related schemes that require state-dependent feedback operations, erasure is a state-independent operation, so readout is not intrinsically necessary. Accordingly, we consider that there is no need to account for the readout cost in our analysis.
 
On the other hand, the charge sensor generates Joule heat, which could potentially raise the temperature of the SC. However, the difference between the sample stage temperature (300 K) and the measured value obtained by Gaussian fitting (305 K) can be explained by the junction capacitance $C_\mathrm{j}$ (see Supplemental Material Section III), suggesting that the effect of Joule heating is negligibly small.
     
\end{document}